\documentclass[runningheads]{cl2emult}
\usepackage{psfig}
\usepackage{makeidx}  
\usepackage{subeqnar} 
\usepackage{multicol} 
\usepackage{amsmath}
\usepackage{cropmark} 
\usepackage{engi}  
\begin{document}
\bibliographystyle{unsrt}

\title*{\vspace*{-2.0cm}{\footnotesize In: In K.~Popp and W.~Schiehlen (eds.), {\em System Dynamics of
\vspace{-0.3cm}\\
  Long-Term Behaviour of Railway Vehicles, Track and Subgrade},
\vspace{-0.3cm}\\
 Lecture Notes
  in Applied Mechanics, Springer (Berlin, 2002).}
\vspace{0.4cm}\\
Rigid body dynamics of railway ballast}
\titlerunning{Rigid body dynamics of railway ballast}
\author{Thomas Schwager and Thorsten P\"oschel}
\authorrunning{T. Schwager \& T. P\"oschel}
\institute{Humboldt-Universit\"at zu Berlin, Charit\'e, Institut f\"ur Biochemie,\\ Monbijoustra{\ss}e 2, D-10117 Berlin, Germany}

\maketitle
\begin{abstract}
A method for the discrete particle simulation of of almost rigid, sharply edged frictional particles, such as railway ballast is proposed. In difference to Molecular Dynamics algorithms, the method does not require knowledge about the deformation-force law of the material. Moreover, the method does not suffer from numerical instability which is encountered in MD simulations of very stiff particles.
\end{abstract}

\section{Introduction}
Traditionally, the subgrade of railway tracks is modeled using continuum mechanics. These methods have been proven to yield reliable results in many applications and have been developed to standard methods. In certain applications, however, continuum models fail in describing the mechanical properties of the subgrade. This is the case when the ballast must not be considered as a continuous medium, but when the granularity of the material is important. Typical processes which cannot be sufficiently explained by continuum models are sedimentation of the ballast due to recurrent load, abrasion of the ballast particles which leads to less efficient damping properties and the formation of force chains inside the ballast material. 

In the past decade mainly by physicists much work has been done in the field of Molecular Dynamics of granular material, i.e., the numerical simulation of granular material as many-particle systems. This technique was applied to many interesting systems and has contributed to the explanation of several exciting and technologically important effects, such as mixing and demixing of granular materials, avalanche statistics on sand heaps, milling processes, convection dynamics in shaken granular materials and others. Many examples of such simulations can be found, e.g., in \cite{HerrmannHoviLuding:1998}.

The idea of Molecular Dynamics is to simulate the granular material as a many particle system and to determine the dynamics of the system by numerical integration of Newton's equation of motion for each of the $N$ particles:
\begin{equation}
  \label{eq:Newton}
  \begin{split}
  \ddot{\vec{r}}_i =&\vec{F}_i/ m_i\,\\
  \ddot{\phi}_i =& {\hat{J}_i}^{-1} \vec{M}_i \,,
\end{split}
\end{equation}
where $\vec{r}_i$ and $\vec{\phi}_i$ are the position and the orientation of the $i$-th particle of mass $m_i$ and moment of inertia $\hat{J}_i$  while $\vec{F}_i$ and $\vec{M}_i$ are the force and the torque acting on this grain. In three dimensions Eqs. (\ref{eq:Newton}) establish a set of $6N$ coupled non-linear differential equations which have to be numerically integrated. 
The force $\vec{F}_i$ consists of gravity and of the interaction force of the particle $i$ with other particles $j$
\begin{equation}
  \label{eq:f}
  \vec{F}_i=m_i \vec{g} + \sum_j \vec{F}_{ij}\,,
\end{equation}
where $\vec{g}$ is the gravitational acceleration. Granular particles interact with each other only if they are in mechanical contact. For spheres of radii $R_i$ and $R_j$ we write
\begin{equation}
  \label{eq:FF}
  \vec{F}_{ij}=\left\{
    \begin{tabular}{ll}
$F^N_{ij} \vec{n}_{ij} + F^T\vec{t}$ ~~& if~~$\left|\vec{r}_i-\vec{r}_j\right|<R_i+R_j$\\
0 & else\,, 
    \end{tabular}
\right.
\end{equation}
where $F^N$ and $F^T$ are the components of the force in normal and tangential direction with respect to the inter-center vector $\vec{r}_i-\vec{r}_j$ and $\vec{n}$ and $\vec{t}$ are the corresponding unit vectors. There exist several models for the interaction forces in normal and tangential direction $F^{N/T}_{ij}\left(\vec{r}_i,\vec{r}_j, \dot{\vec{r}}_i, \dot{\vec{r}}_j\right)$ (see, e.g. \cite{SchaeferDippelWolf:1995}) which shall not be discussed in detail here. 

\section{Molecular Dynamics fails for the simulation of railway ballast}\label{sec:MDunsutable}

There are many examples where granular systems have been simulated by Molecular Dynamics, however, except for few examples, realistic simulations have been achieved only for systems where the {\em dynamical} behavior of the grains dominates the system properties. When the static properties of the particle system become important, i.e., when the relative velocities are small or zero, the details of the interaction force become essential for understanding the system behavior. We are faced with two major problems:
\begin{itemize}
\item The interaction force of contacting particles must be known as a function of the particle positions and velocities. In the case of sharply edged grains such as railway ballast, this function is unknown.
\item As soon as the realistic simulation of static properties matters for the system behavior,  the simulation slows down extremely which implies that to achieve affordable computation times one has to make simplifying assumptions on the particle contact which are not justified from the point of view of mechanics and material science. 
\end{itemize}
For these reasons we believe that the described Molecular Dynamics method is {\em in principle} unsuitable for the simulation of railway ballast. Below we list arguments which support this statement:
\begin{enumerate}
\item {\bf Elastic normal force:} Whereas the normal elastic force of contacting spheres is given by the Hertz law
  \begin{equation}
    \label{eq:Hertz}
    F^{N,{\rm el}}_{ij}=\frac23\frac{Y\sqrt{R_{\rm eff}}}{1-\nu^2}~\xi^{3/2}_{ij}
  \end{equation}
with $Y$, $\nu$, and $R_{\rm eff}$ being the Young modulus, the Poisson ratio, and the effective radius, respectively, this force is not known for more complicatedly shaped particles. For smooth particles (when the local radius of the contact area is large as compared to the compression $\xi_{ij}\equiv R_i+R_j-\left|\vec{r}_i-\vec{r}_j\right|$)  the function (\ref{eq:Hertz}) is certainly a good approximation, for sharply edged particles, however, this function fails. Some authors assume that the normal force is proportional to what they call ``overlap'', i.e., the volume of the compressed material or (yet more simple) in two dimensions the ``compression length'' (for spheres the value $\xi$) which is certainly incorrect even for the most simple case of contacting spheres and the more for more complicatedly shaped grains. 

\item {\bf Dissipative normal force:} The dissipative normal force of contacting bodies is unknown, in general. For viscoelastically colliding spheres and other smooth contacting surfaces it is given by 
  \begin{equation}
    \label{eq:diss}
    F^{N,{\rm diss}}_{ij}=A\dot{\xi}\frac{\partial}{\partial \xi} F^{N,{\rm el}}\,.    
  \end{equation}
The prefactor $A$ is a complicated function (for details see \cite{BrilliantovSpahnHertzschPoeschel:1994}) which contains the viscous constants of the material which are unknown in general. For particles which are not smooth, such as railway ballast, even the functional form of the dissipative force is unknown. The frequently applied force law $F^{N,{\rm diss}}_{ij}\propto \dot{\xi}$ is not justified and fails even for spheres.

\item {\bf Tangential force:} Whereas for smooth bodies the normal force can be determined from bulk properties of the material, the tangential force is determined by the bulk and by surface properties. A natural (phenomenological) assumption is 
  \begin{equation}
    \label{eq:coulomb}
    F^{T}_{ij}\le\mu     F^{N}_{ij}\,,
  \end{equation}
where $\mu$ is the Coulomb friction parameter. Unfortunately, this model is not sufficient for static systems: Assume a (non-spherical) particle which rests on an inclined plane (angle $\alpha$). Its normal force is $F^N=mg\cos\alpha$ and there is a corresponding tangential force too. To prevent the particle from sliding along the plane one has to assume an additional force which mimics static friction, e.g. \cite{SchaeferDippelWolf:1995}. This force cannot be derived from material or surface properties, hence, it is arbitrary. Choices for this function which can be found in literatures are even in disagreement with basic mechanics.

\item {\bf Numerical integration of Newton's law:} Besides the fact that the numerical integration of the set (\ref{eq:Newton}) for a relevant number of particles $N$, e.g., $N=1,000$ which establish a quite small system of $10\times 10 \times 10$ particles in three dimensions,  requires serious numerical effort, there are principal problems when applying Molecular Dynamics to a particle system:
  \begin{itemize}
  \item For very rigid particles the gradient of the force is large and, therefore, requires a very small integration time step. The more rigid the particle material the slower progresses the simulation. Assuming a time step of $\Delta t=10^{-7} $ sec which is used in many MD simulations of granular material, one needs $10^9$ integration steps of the equations of motion (\ref{eq:Newton}) to achieve a real time of $100 $ sec only. Hence, a simulation of long time behavior using MD seems to be unrealistic.
  \item The accuracy of the algorithm cannot be increased too much by reducing the time step. Frequently, predictor-corrector algorithms are used, which require powers of the time step, $(\Delta t)^2=10^{-14}$, $(\Delta t)^3=10^{-21}$, etc. (again for $\Delta t=10^{-7}$). If we ignore the prefactor of this powers in the integration scheme to obtain a crude estimate, one has to sum numbers of the orders ${\cal} O(1)$, ${\cal O}\left(10^{-7}\right)$, ${\cal O}\left(10^{-14}\right)$, etc. Double precision numbers have a mantissa of 15 digits, therefore, one cannot sum numbers which are different by more than $10^{15}$. This limits the minimal value of the integration time step. If one requires higher precision one needs real numbers in quadruple precision or higher which implies higher memory consumption and (more importantly) yet larger computation time since multiplication of two numbers in quadruple precision requires approximately quadruple time as a multiplication of double precision numbers.

We are aware that these numbers are a very crude estimate which have been given to illustrate the problem. Of course, more sophisticated integration schemes suffer less critical from the addition of different numbers. In principle, however, the problem persists.
  \end{itemize}

\end{enumerate}

\section{Rigid Body Dynamics}
\label{sec:starr.prinzip}

In Molecular Dynamics simulations the trajectories of the particles are determined by numerical integration of Newton's equations of motion. As discussed above, this implies the ``soft particle assumption'', i.e., the particle deform under load. Since the particles are very hard but not completely rigid we encounter serious numerical problems as described above.

\begin{figure}[htbp]
  \centerline{\psfig{figure=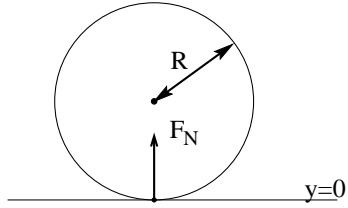,width=4.5cm}}
  \caption{A rigid sphere resting on a rigid flat surface.}
  \label{fig:simpleexample}
\end{figure}

The Rigid Body Dynamics originates from the opposite idea: The interaction forces are determined from the required behavior of the particles. This method is, therefore, suited to simulate perfectly rigid particles without the necessity to specify a certain force-deformation law. We consider the example of a rigid sphere which rests on a rigid flat surface (see Fig. \ref{fig:simpleexample}).
There is a point-like contact of the sphere and the surface, hence, there is a contact force $F^N$ in vertical direction. Moreover gravity acts on the sphere causing a force $-mg$. If one would chose the contact force $F^N=0$, the sphere would move downwards with an acceleration $g$, i.e., it would penetrate the surface. This unphysical behavior has to be avoided by the proper choice of the force $F^N$ due to the following conditions:
\begin{enumerate}
\item {\em Contact forces have to be chosen in order to avoid penetration of contacting particles.}\\ For our example $F^N\ge mg$ follows. On the other hand, a force $F^N>mg$ would lead to unphysical behavior since the sphere would move upwards. This is avoided by the second condition.

\item {\em A contact force vanishes when the contact breaks.}\\ A contact is said to break if the normal component of the relative acceleration of the concerned particles, or their normal velocity is larger than zero. (The relative velocity is counted positive if the particles separate from each other.)

\item {\em There are no attractive normal forces.}\\
In our example $F^N>mg$ causes the contact to break which implies that the contact force vanishes. Therefore, from these conditions follows $F^N=mg$. For this choice the total force is zero and the sphere rests on the plane.
The conditions 1.-3. are sufficient to describe any particle system as long as there are no friction forces. For systems with friction we need one more condition:

\item {\em Friction forces act in parallel with the contact plane, i.e., perpendicular to $F^N$. Given the tangential force $F^*$ which is necessary to keep two particles from sliding. Then the acting tangential force is $\left|F\right|=\min\left(\left|F^*\right|,\left|\mu F^N\right|\right)$. Its sign has to be chosen opposite to the tangential relative acceleration (or the tangential relative velocity).}\\ In agreement with Coulomb's friction law the particles slide only if $\left|F^T\right| \ge \left|F^N\right|$. If this condition is fulfilled the friction force adopts its maximal value $\pm \mu F^N$.
\end{enumerate}
To perform simulations one has to derive the forces of the particles in normal and tangential directions from these four conditions. The corresponding algorithm will be explained in sections \ref{sec:mathematical} and \ref{sec:Dantzig}.

In our simple example (Fig. \ref{fig:simpleexample}) there exists only one contact between the sphere and the plane. In more complex situations, e.g. for a resting cube, there are contacts areas instead of points. These contacts may be always reduced to point contacts. It will be shown that the described conditions are sufficient to determine the forces and torques which act on the particles, provided there are not too many contacts in the system. If the number of contacts is too large only the total force and the total torque which act on a particle may be determined, but not each of the pairwise contact forces. For the computation of the particle trajectories, however, the total forces and torques are sufficient. We will return to this issue in Sec. \ref{sec:indeterminacy}.
\medskip

Rigid Body Dynamics has been intensively studied in the past two decades. Descriptions of the algorithm can be found in \cite{Loetstedt:1982,Loetstedt:1981}. The core of the algorithm is the numerical computation of the contact forces which is a Linear Complementarity Problem \cite{CottleDantzig:1968,CottlePangStone:1992}. An efficient algorithm for this type of problems can be found in \cite{Baraff:1989,Baraff:1991,Baraff:1994,Baraff:1996}. Rigid Body Dynamics has also been applied to granular systems, e.g. \cite{Moreau:1994,Roux:2000}, where frictionless smooth spheres have been simulated. Systems of granular particles subject to friction have been studied, e.g., in \cite{Jean:1995}.

Due to our understanding the Rigid Body Dynamics is much better suited for the simulation of railway ballast for the following reasons:
\begin{itemize}
\item Ballast particles are irregularly shaped and sharply edged. Even if the bulk material properties were precisely known, the contact force law is unknown due to the complicated shape. 
\item Ballast particles are very stiff which implies that the gradient of the interaction force is very steep. In this regime the numerical integration of Newton's equation is problematic. For the dynamics of the system the deformation of single particles is unimportant, i.e., the Rigid Body assumption is well justified.
\item  Static friction, whose treatment in MD-simulations is problematic too, is essential for the dynamics of the system. It is correctly modeled in Rigid Body Dynamics. 
\item  The long time behavior of railway ballast is affected by abrasion and fragmentation of particles. Molecular Dynamics of fragmenting particles may cause artifacts for several reasons which cannot be discussed here in detail (see \cite{PoeschelSchwager:2002Algo}). Rigid Body Dynamics is very well suited for this case.
\end{itemize}

\section{Schedule of Rigid Body Simulations}
\label{sec:ablauf}

The state of the granular system is described by the position and orientation of its particles and by the according time derivatives. Contacts between the particles may be classified as sticking and sliding contacts. In the due of time the contact network is modified by creation and termination of contacts as well as by transformation of sticking contacts into sliding ones and vice versa. Whenever the contact network is modified the state of the system changes qualitatively. The simulation proceeds in discrete time steps. Each of them consists of
\begin{enumerate}
\item {\bf Contact detection:} all existing contacts are registered.

\item {\bf Treatment of collisions:} A collision takes place if two contacting particles move with negative normal relative velocity. In this case one cannot determine a {\em finite} contact force which would avoid penetration of the particles since {\em any} force, however large it is, would need a short but finite path to retard the colliding particles. Hence, mutual penetration would be unavoidable. Therefore, we need a special treatment for collisions (see Sec. \ref{sec:collisions}).

\item {\bf Formulation of the geometry equations:} After a collision, in general, there is a number of contacts of particles which have a positive normal relative velocity, i.e., the particles lose contact. These contacts have to be erased from the list of contacts. The normal components of the relative velocities at all remaining contacts are zero. We have to establish the geometry equations which contain the information about the geometry of the system (see Sec. \ref{sec:mathematical}).

\item {\bf Computation of the forces:} Section \ref{sec:Dantzig} deals with the computation of the relative accelerations by means of the geometry equations.

\item {\bf Integration of the equations of motion:} Finally we have to integrate the equations of motion for all particles. During this operation it may be necessary to update the geometry equations and to repeat the computation of the forces according to the integration scheme used. 
\end{enumerate}

\section{Mathematical description}\label{sec:mathematical}
\begin{figure}[htbp]
  \centerline{\psfig{figure=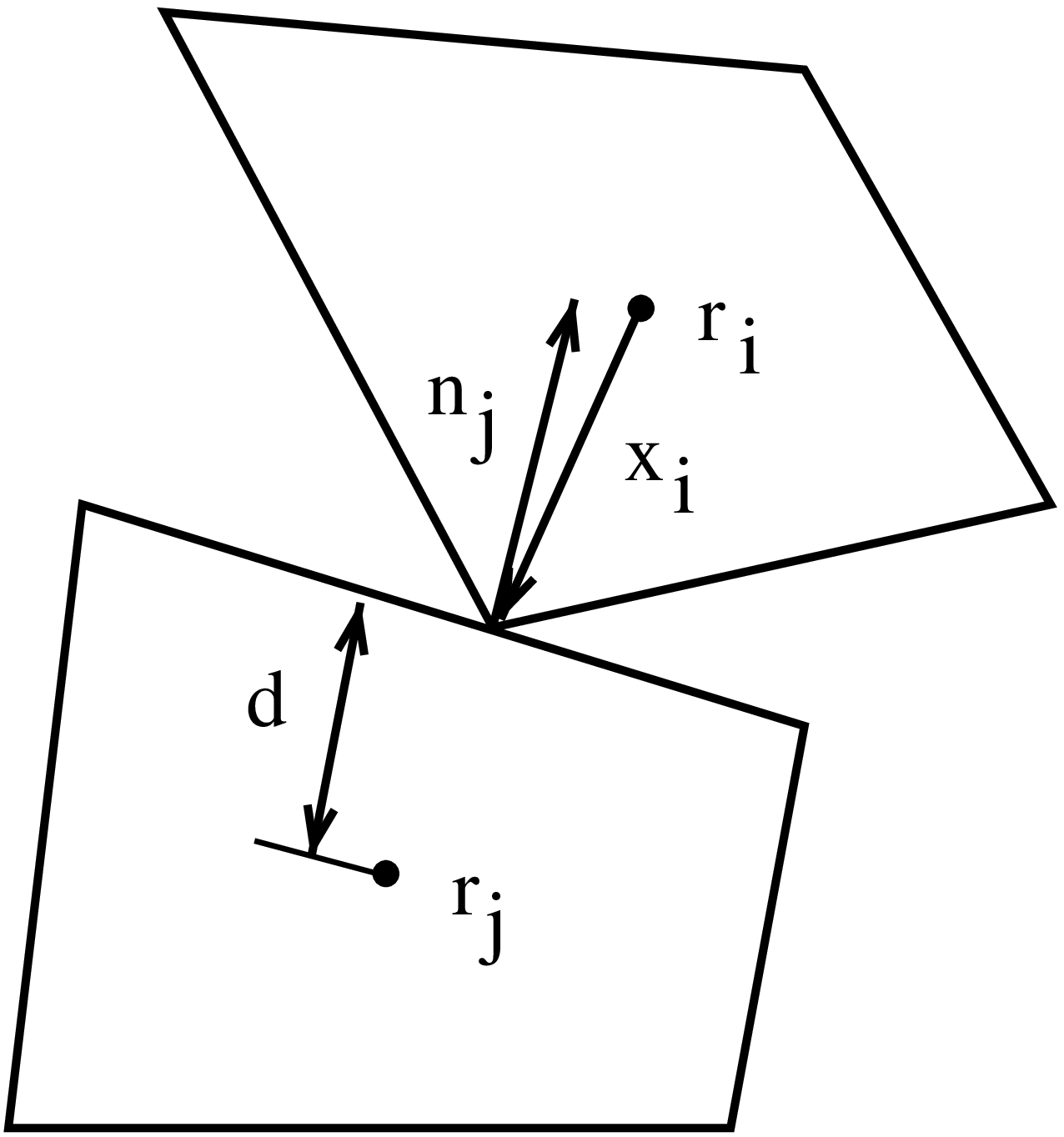,width=4cm}\hspace{0.9cm}\psfig{figure=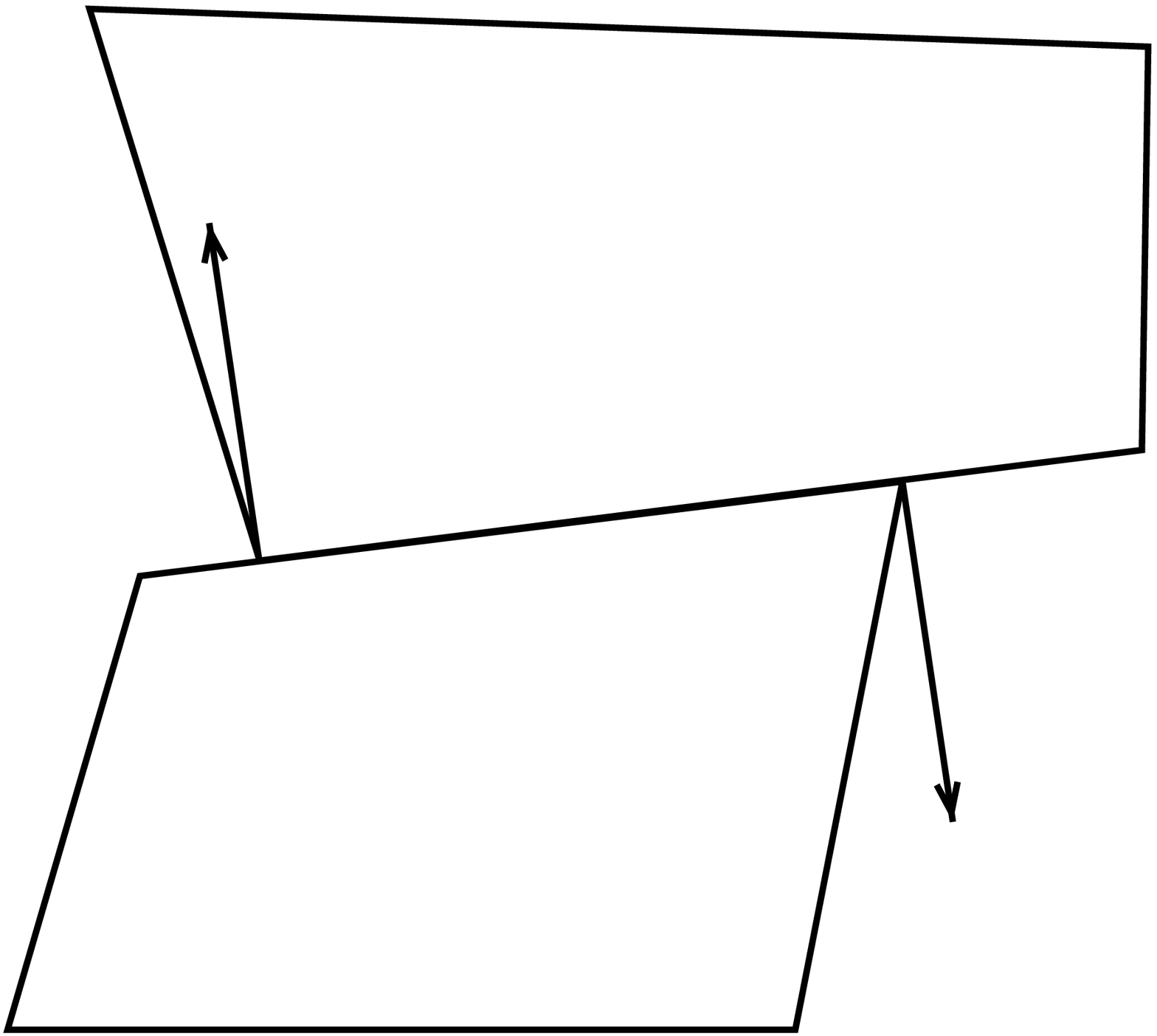,width=4cm}}
  \caption{Left: a face-edge contact. $\vec{r}_i$ and $\vec{r}_j$ are the center of mass positions, $\vec{x}_i$ is the coordinate of the contacting edge relative to the center of mass of particle $i$ and $\vec{n}_j$ is the normal of the contacting face of particle $j$. Right: face-face contacts can be reduced to two face-edge contacts. The face normals of each of the contacts are displayed as well.}
  \label{fig:sketch}
\end{figure}

For convenience at first we will restrict to frictionless particles. When the mathematical framework has been developed for this simplified case we will then introduce friction forces between particles. 
The rigidity of the particles is enforced by means of mathematical motion constraints of the form 
\begin{equation}
  g(q)\ge0~,\label{eq:constraint}
\end{equation}
where $q$ is a vector containing the positions (center of mass position and orientation) of all particles of the system. The constraint function $g$ shall be zero if particles in the system are in contact and larger than zero otherwise. For spheres, which is the most simple case, the constraint function reads
\begin{equation}
  g(q)=\left|\vec{r}_i-\vec{r}_j\right|-R_i-R_j~.
\end{equation}
If the spheres would deform each other ($\left|\vec{r}_i-\vec{r}_j\right|<R_i+R_j$) the function $g(q)$ would be negative, if the particles touch each other it would be zero. To prevent the deformation we require $g(q)$ to be positive or zero, i.e., it has to fulfill the condition (\ref{eq:constraint}). For sharp edged particles, such as particles which are described by polyhedrons or polygons, the motion constraints are
\begin{equation}
  g(q)=\vec{n}_j\left(\vec{r}_i+\vec{x}_i-\vec{r}_j\right)-d~,\label{eq:normalconstraint}
\end{equation}
where $\vec{n}_j$ is normal of the face of particle $j$ which is in contact with an edge of the particle $i$. The vectors $\vec{r}_i$ and $\vec{r}_j$ are the center of mass positions of the particles, the edge of particle $i$ that is in contact with the particle $j$ is at position $\vec{r}_i+\vec{x}_i$. The constant $d$ is the distance of the contacting face of $j$ from the center of the particle. The left picture in Fig. \ref{fig:sketch} shows a sketch of a typical contact of two particles. Face-face contacts can be described by two face-edge contacts (see Fig. \ref{fig:sketch}).

Every motion constraint corresponds to a scalar contact force $f$. According to d'Alembert's principle the direction of the contact force is given by the spacial derivative of $g$ with respect to all components of the coordinate vector $q$, namely $\partial g/\partial q$. The contact force that acts on a certain particle $i$ is $f\partial g/\partial q_i$, with $q_i$ being the coordinates of particle $i$. We can formulate the equation of motion for the particles\footnote{Particle indices are written in Latin letters, contact indices in Greek letters.}
\begin{eqnarray}
  \hat{M}_i\ddot{q}_i&=&Q_i+\sum\limits_{\alpha}f_\alpha\frac{\partial g_\alpha}{\partial q_i}\label{eq:eqnofmotion}\\
  g_\alpha(q)&\ge&0~.
\end{eqnarray}
$\hat{M}_i$ is the mass matrix of the particle $i$, which has the form
\begin{equation}
  \hat{M}_i=\left(
    \begin{array}{rrrr}
      m_i&0&0&0\\
      0&m_i&0&0\\
      0&0&m_i&0\\
      0&0&0&\hat{J}_i
    \end{array}\right)~,
\end{equation}
where $\hat{J}_i$ is the moment of inertia tensor. In two dimensional systems $\hat{J}_i$ is only a scalar $J_i$ and there are only 2 entries of $m_i$. $Q_i$ finally is the external force (and torque) acting on particle $i$. This is usually gravitation, but other external forces can be incorporated at this point as well. 

Although we formulated the motion constraints in the form $g(q)\ge0$ to allow separation of particles, contact forces can only act if particles are actually in contact. Therefore, constraints which are strictly positive, i.e. $g(q)>0$ (the particles are separated), can be disregarded. These constraints are said to be inactive. The remaining constraints, the active ones, are thus satisfied by $g(q)=0$. Since the $g(q)$ have to remain non-negative their time derivatives have to be non-negative as well. We therefore have
\begin{eqnarray}
  \dot{g}_\alpha&=&\frac{\partial g_\alpha}{\partial q_k}\dot{q}_k\ge0\\
  \ddot{g}_\alpha&=&\frac{\partial g_\alpha}{\partial q_k}\ddot{q}_k+\frac{\partial^2g_\alpha}{\partial q_k\partial q_l}\dot{q}_k\dot{q}_l\ge0~.\label{eq:ddottgalpha}
\end{eqnarray}
For simplicity of notation we used the Einstein convention, i.e. summation over doubly occurring indices $k$ and $l$ is implied. These time derivatives are the relative velocity and relative acceleration of the particles at their contact points. It is important to note that $\dot{g}_\alpha$ and $\ddot{g}_\alpha$ are not the relative velocity or acceleration of the centers of mass of the particles but of the points of both particles which are actually in contact. It can easily happen that the relative velocity or acceleration of the contact points are positive (the particles are about to separate) although their centers of mass approach each other. 

We insert the equation of motion (\ref{eq:eqnofmotion}) into (\ref{eq:ddottgalpha}) and find
\begin{eqnarray}
  \ddot{g}_\alpha&=&\frac{\partial g_\alpha}{\partial q_k}\left[\hat{M}_k^{-1}\left(Q_k+\sum\limits_\beta f_\beta\frac{\partial g_\beta}{\partial q_k}\right)\right]+\frac{\partial^2g_\alpha}{\partial q_k\partial q_l}\dot{q}_k\dot{q}_l\nonumber\\
  &=&\frac{\partial g_\alpha}{\partial q_k}\left(\hat{M}_k^{-1}Q_k\right)+\frac{\partial^2g_\alpha}{\partial q_k\partial q_l}\dot{q}_k\dot{q}_l+\frac{\partial g_\alpha}{\partial q_k}\hat{M}_k^{-1}\left(\sum\limits_\beta f_\beta\frac{\partial g_\beta}{\partial q_k}\right)~.
\end{eqnarray}
The first term on the right hand side describes the action of the external forces, the second term describes the action of inertial forces as, e.g., centrifugal force and Coriolis force, the third term finally describes the action of the contact forces. 
We can rewrite this equation as
\begin{equation}
  \ddot{g}_\alpha=b_\alpha+\sum\limits_\beta A_{\alpha\beta}f_\beta~,\label{eq:geometryeq1}
\end{equation}
with $A_{\alpha\beta}$ and $b_\alpha$ abbreviating
\begin{equation}
  \begin{split}
    A_{\alpha\beta}&=\frac{\partial g_\alpha}{\partial q_k}\hat{M}_k^{-1}\frac{\partial g_\beta}{\partial q_k}\\
    b_\alpha&=\frac{\partial g_\alpha}{\partial q_k}\left(\hat{M}_k^{-1}Q_k\right)+\frac{\partial^2g_\alpha}{\partial q_k\partial q_l}\dot{q}_k\dot{q}_l~.
  \end{split}
\end{equation}
From now on we will denote the relative acceleration of the contacting particles at their contact points -- the contact acceleration -- as $a_\alpha$ instead of $\ddot{g}_\alpha$. Equation (\ref{eq:geometryeq1}) turns into 
\begin{equation}
  a_\alpha=b_\alpha+A_{\alpha\beta}f_\beta~,\label{eq:geometryeq2}
\end{equation}
where again summation over $\beta$ is implied. We will call this equation the geometry equation. By means of this equation and the consistency conditions introduced in Sec. \ref{sec:starr.prinzip} we can now determine the contact forces $f_\beta$. The consistency conditions read
\begin{equation}
  \begin{split}
    a_\alpha&\ge0\\
    f_\alpha&\ge0\\
    a_\alpha f_\alpha&=0~.
  \end{split}
  \label{eq:conditions3}
\end{equation}
The first condition prevents deformation of particles, the second one excludes attractive forces and the third one requests that contact forces may only act if the particles stay in contact, i.e., if $a_\alpha=0$. These conditions together with the geometry equation (\ref{eq:geometryeq2}) allows to determine the unknown contact forces $f_\alpha$. The whole system consisting of Eq. (\ref{eq:geometryeq2}) and the conditions (\ref{eq:conditions3}) is called a Linear Complementarity Problem. It can be solved by Dantzig's algorithm \cite{CottleDantzig:1968}.

To incorporate friction we introduce additional motion constraints which shall, if possible, impede a tangential motion of the contacting particles. For polygonal particles they are of the form
\begin{equation}
  g(q)=\vec{t}_j\left(\vec{r}_i+\vec{x}_i-\vec{r}_j-\vec{x}_j\right)~.
\end{equation}
This constraint has a similar form as the constraint of the normal motion of the particles (\ref{eq:normalconstraint}) but instead of the normal unit vector of the contacting face of particle $j$ the tangential unit vector $\vec{t}_j$ appears, thus ensuring that the edge of particle $i$ at position $\vec{r}_i+\vec{x}_i$ does not move along the face of particle $i$ away from the point $\vec{r}_j+\vec{x}_j$ of initial contact. 

These motion constraints are, however, of different nature than the normal motion constraints. Whereas in the case of the normal motion the constraints must never be violated, the constraints on the tangential motion may actually be violated, as it happens when the particles start to slide. This is due to the fact that the magnitude of the friction forces are limited by $\mu f_N$, with $\mu$ being the friction constant and $f_N$ the corresponding normal force. As reflected by the consistency condition 4 (ref. Sec. \ref{sec:starr.prinzip}) the friction force must adopt its maximum value if particles actually slide, i.e. if the tangential motion constraint is inactive. Thus, in this case the value of the friction force is determined without need of further consideration. In contrast to the case of normal motion constraints we may not neglect the inactive constraints because now the corresponding contact forces are non-zero and thus the constraint has to be kept in consideration in order to determine the direction of the tangential force. 

Since friction causes only further motion constraints there is, in principle, no need of further discussion of the problem. The geometry equation (\ref{eq:geometryeq2}) can describe systems with friction as well. For simplicity of notation it is worth, however, to consider normal and friction forces and their corresponding motion constraints separately. We now have 2 classes of motion constraints
\begin{equation}
  \begin{split}
    g^N(q)\ge&0\\
    g^T(q)=&0    
  \end{split}
\end{equation}
and the corresponding contact forces $f^N$ and $f^T$. Now the equation of motion reads 
\begin{equation}
  \hat{M}_k\ddot{q}_k=Q_k+\sum\limits_\alpha\left(f^N_\alpha\frac{\partial g^N_\alpha}{\partial q_k}+f^T_\alpha\frac{\partial g^T_\alpha}{\partial q_k}\right)~.
\end{equation}
For the second time derivative of the motion constraints we obtain
\begin{equation}
  \begin{split}
    \ddot{g}_\alpha^N=&\frac{\partial g^N_\alpha}{\partial q_k}\left(\hat{M}_k^{-1}Q_k\right)+\frac{\partial^2g^N_\alpha}{\partial q_k\partial q_l}\dot{q}_k\dot{q}_l+\frac{\partial g^N_\alpha}{\partial q_k}\hat{M}_k^{-1}\left(\sum\limits_\beta f^N_\beta\frac{\partial g^N_\beta}{\partial q_k} + f^T_\beta\frac{\partial g^T_\beta}{\partial q_k}\right)\\
    \ddot{g}_\alpha^T=&\frac{\partial g^T_\alpha}{\partial q_k}\left(\hat{M}_k^{-1}Q_k\right)+\frac{\partial^2g^T_\alpha}{\partial q_k\partial q_l}\dot{q}_k\dot{q}_l+\frac{\partial g^T_\alpha}{\partial q_k}\hat{M}_k^{-1}\left(\sum\limits_\beta f^N_\beta\frac{\partial g^N_\beta}{\partial q_k} + f^T_\beta\frac{\partial g^T_\beta}{\partial q_k}\right)\,.    
  \end{split}
\end{equation}
Renaming again $\ddot{g}_\alpha^{\{N,T\}}$ by $a_\alpha^{\{N,T\}}$ and using the abbreviations
\begin{equation}
  \begin{split}
    b_\alpha^N=&\frac{\partial g^N_\alpha}{\partial q_k}\left(\hat{M}_k^{-1}Q_k\right)+\frac{\partial^2g^N_\alpha}{\partial q_k\partial q_l}\dot{q}_k\dot{q}_l\\
    b_\alpha^T=&\frac{\partial g^T_\alpha}{\partial q_k}\left(\hat{M}_k^{-1}Q_k\right)+\frac{\partial^2g^T_\alpha}{\partial q_k\partial q_l}\dot{q}_k\dot{q}_l
  \end{split}
\end{equation}
and 
\begin{equation}
  \begin{tabular}{llll}
    $\displaystyle{A_{\alpha\beta}^{NN}}$&$\displaystyle{=\frac{\partial g^N_\alpha}{\partial q_k}\hat{M}_k^{-1}\frac{\partial g^N_\beta}{\partial q_k}}$~~~~~~~&$\displaystyle{A_{\alpha\beta}^{NT}}$&$\displaystyle{=\frac{\partial g^N_\alpha}{\partial q_k}\hat{M}_k^{-1}\frac{\partial g^T_\beta}{\partial q_k}}$\\
    $\displaystyle{A_{\alpha\beta}^{TN}}$&$\displaystyle{=\frac{\partial g^T_\alpha}{\partial q_k}\hat{M}_k^{-1}\frac{\partial g^N_\beta}{\partial q_k}}$~~~~~~~&$\displaystyle{A_{\alpha\beta}^{TT}}$&$\displaystyle{=\frac{\partial g^T_\alpha}{\partial q_k}\hat{M}_k^{-1}\frac{\partial g^T_\beta}{\partial q_k}}$
  \end{tabular}
\end{equation}
we can write the modified geometry equations
\begin{equation}
  \label{eq:geometrymodified}
  \begin{split}
    a_\alpha^N=&b_\alpha^N+\sum\limits_\beta\left(A_{\alpha\beta}^{NN}f_\beta^N + A_{\alpha\beta}^{NT}f_\beta^T\right)\\
    a_\alpha^T=&b_\alpha^T+\sum\limits_\beta\left(A_{\alpha\beta}^{TN}f_\beta^N + A_{\alpha\beta}^{TT}f_\beta^T\right)~.
  \end{split}
\end{equation}
The full set of consistency conditions then reads
\begin{equation}
  \label{eq:conditions5}
  \begin{split}
    a_\alpha^N\ge&0\\
    f_\alpha^N\ge&0\\
    a_\alpha^Nf_\alpha^N=&0\\
    \left|f_\alpha^T\right|\le&\mu f_\alpha^N\\
    a_\alpha^T\left(\left|f_\alpha^T\right|-\mu f_\alpha^N\right)=&0~.
  \end{split}
\end{equation}
Equations (\ref{eq:geometrymodified}) together with the conditions (\ref{eq:conditions5}) can be solved with a modified Dantzig's algorithm \cite{Baraff:1994} which will be discussed in the next section. Note that some of the tangential forces may be directly determined by the respective normal forces, i.e., when sliding at this contact occurs. The consistency conditions for these forces have to be fulfilled, nevertheless. 

\section{Dantzig's Algorithm}\label{sec:Dantzig}

Contacts can be classified into breaking contacts ($a^N=0$ thus $f^N=f^T=0$), permanent static contacts, ($a^N=0$ and $a^T=0$) or permanent sliding contacts ($a^N=0$ but $a^T\ne0$). If we knew \`a priory into which category each contact belongs the contact forces could be determined by solving an inhomogeneous system of linear equations which consists of all equations for which either $a_\alpha^N=0$ or $a_\alpha^T=0$, with the corresponding $f_\alpha^N$ and $f_\alpha^T$ as variables. All remaining normal forces are zero, the remaining tangential forces assume their maximum values. 
Unfortunately the contact classification is only known if we know the contact forces as well.

We apply Dantzig's Algorithm to determine the forces together along with the corresponding contact classification. It starts with considering a certain contact, disregarding all others, i.e., their contact forces are set to zero. After having found a solution for this contact its classification is also known. The algorithm proceeds then with the next contact. Again the contact forces are determined preserving the consistency of all contacts considered before. In this process the contact classifications of the already consistent contacts may be changed if necessary. The process is repeated until the last contact has been classified. 

All contacts are assigned to one of four lists:
\begin{itemize}
\item List $NC$ of breaking contacts
\item List $C_F$ of permanent static contacts
\item List $C^{\pm}$ of permanent sliding contacts. In list $C^+$ are all contacts where $f^T=\mu f^N$, in $C^-$ all contacts where $f^T=-\mu f^N$.
\end{itemize}
All contacts in the above lists are considered to be consistent, i.e., they satisfy the consistency conditions. The classification is done successively for all contacts $\alpha$ by:
\begin{enumerate}
\item Check if the normal force $f_\alpha^N=0$ satisfies the consistency conditions $a_\alpha^N\ge0$. If this is the case the contact is consistent and belongs to $NC$. 
\item If $a_\alpha^N<0$ we have to increase the normal contact force $f_\alpha^N$ to obtain a non-negative normal acceleration. However, increasing the normal force $f_\alpha^N$ will change the contact accelerations of the already classified contacts as well. Since for the permanent contacts $\beta$ we need $a_\beta^N=0$ (and for the static contacts also $a_\beta^T=0$) this would invalidate the classification of these contacts. To preserve the consistency of the classified contacts we will have to change the contact forces of the the permanent contacts as well. We now have to determine how much we have to change these contact forces for a given increase $s$ of the new force $f_\alpha^N$ to keep $a_\beta^N=0$ and, if necessary $a_\beta^T=0$. To calculate the necessary changes we formulate a reduced set of geometry equations:
\begin{equation}
  \label{eq:geometryreduced}
  \begin{split}
    0=a_\beta^N=&b_\beta^N + A^{NN,\rm red}_{\beta\gamma}f_\gamma^N + A^{NT,\rm red}_{\beta\gamma}f_\gamma^T + A^{NN}_{\beta\alpha}f_\alpha^N\\
    0=a_\beta^T=&b_\beta^T + A^{TN,\rm red}_{\beta\gamma}f_\gamma^N + A^{TT,\rm red}_{\beta\gamma}f_\gamma^T + A^{TN}_{\beta\alpha}f_\alpha^N\,.
  \end{split}
\end{equation}
The reduced set of geometry equations can be obtained from the original geometry equations (\ref{eq:geometrymodified}) disregarding all breaking contacts and replacing $f_\beta^T=\pm \mu f_\beta^N$ for all contacts $\beta$ from $C^\pm$ (sliding contacts). Since only permanent contacts remain all contact accelerations in the reduced geometry equations are thus zero. If we now change the new contact force $f_\alpha^N\to f_\alpha^N+s$ we have to vary the previously known contact forces $f_\beta^{\{N,T\}}$ in the reduced geometry equations by an unknown amount $\Delta f_\beta^{\{N,T\}}$ in order to keep the contact accelerations at their required value of zero: 
\begin{equation}
  \begin{split}
    0=&b_\beta^N + A^{NN,\rm red}_{\beta\gamma}\left(f_\gamma^N+\Delta f_\gamma^N\right) + A^{NT,\rm red}_{\beta\gamma}\left(f_\gamma^T+\Delta f_\gamma^T\right)\\
    &+ A^{NN}_{\beta\alpha}\left(f_\alpha^N+s\right)\\
    0=&b_\beta^T + A^{TN,\rm red}_{\beta\gamma}\left(f_\gamma^N+\Delta f_\gamma^N\right) + A^{TT,\rm red}_{\beta\gamma}\left(f_\gamma^T+\Delta f_\gamma^T\right)\\
    &+ A^{TN}_{\beta\alpha}\left(f_\alpha^N+s\right)\,.
  \end{split}
\end{equation}
Inserting Eq. (\ref{eq:geometryreduced}) we find 
\begin{equation}
  \begin{split}
    0=&A^{NN,\rm red}_{\beta\gamma}\Delta f_\gamma^N + A^{NT,\rm red}_{\beta\gamma}\Delta f_\gamma^T + A^{NN}_{\beta\alpha}s\\
    0=&A^{TN,\rm red}_{\beta\gamma}\Delta f_\gamma^N + A^{TT,\rm red}_{\beta\gamma}\Delta f_\gamma^T + A^{TN}_{\beta\alpha}s\,.
  \end{split}
\end{equation}
This is linear system of equations for the unknown $\Delta f_\gamma^{\{N,T\}}$ with the step size $s$ being a parameter. The necessary variations $\Delta f_\gamma^{\{N,T\}}$ are proportional to the step size $s$, hence the solution is of the form
\begin{equation}
  \Delta f_\gamma^{\{N,T\}} = F_\gamma^{\{N,T\}}s~,
\end{equation}
where $F_\gamma^{\{N,T\}}$ is the necessary variation if $s=1$. By inserting the changed values back into the original geometry equations we now know as well by how much the accelerations of the breaking and sliding contacts change. 

\item We increase the value of the new force $f^N_\alpha$ until either the acceleration $a_\alpha^N=0$ (the contact is now consistent) or until the classification of any already consistent contact changes. The classification changes if either
  \begin{enumerate}
  \item a normal acceleration $a_\beta^N>0$ becomes zero: The contact becomes permanent and has to be moved from $NC$ to either $C_F$ or $C^\pm$ according to its present value of the corresponding tangential acceleration. 
  \item a normal force $f_\beta^N>0$ (permanent contact) becomes zero: the contact is now a breaking contact and has to be moved from $C_F$ or $C^\pm$ to $NC$. 
  \item a tangential acceleration $a_\beta^T\ne 0$ becomes zero and the corresponding tangential velocity is zero: the contact now is static and has to be moved from $C^\pm$ to $C_F$. 
  \item a friction force previously of smaller magnitude than its allowed maximum reaches the maximum value $\pm\mu f_\beta^N$: the contact becomes sliding and has to be moved from $C_F$ to $C^\pm$. 
  \end{enumerate}
\item if we have not yet found consistent values for $a_\alpha^N$ and $f_\alpha^N$ we have to proceed with step 2. 
\end{enumerate}

If the contact $\alpha$ is permanent we have to consider the tangential component $a_\alpha^T$ of contact $\alpha$ too. The procedure is very similar to the calculation of $a_\alpha^N$. The only difference is that if $a_\alpha^T>0$ we have to {\em decrease} the friction force until it assumes its negative maximum value and if $a_\alpha^T<0$ we have to increase the friction until it adopts its positive maximum value. 

\section{Collisions}\label{sec:collisions}

In the framework of Rigid Body Dynamics collisions occur if two contacting particles have a negative normal relative velocity at their contact point. We can easily convince ourselves that no finite contact force can prevent a deformation of the particles. No matter how large the force is, it will always take a finite, however small, time to stop the approaching particles, hence they will deform each other. Thus, to prevent deformation of the particles an infinite repulsive force of infinitesimal duration is necessary. It turns out that the total momentum transfer $\Delta p$ between the two particles is finite\footnote{Note that the duration $t_c$ of the collision is only formally the parameter of the limit in the equation above. This limit is actually achieved by starting with deformable particles and increasing their stiffness to infinity. In this limiting process the duration of collision approaches zero.}:
\begin{equation}
  \Delta p = \lim\limits_{t_c\to 0}\int\limits_0^{t_c} fdt={\rm finite}~.
\end{equation}
Resolving multi-particle collisions it turns out that the resulting state of the particles after the collision is not unique. This is due to the infinite stiffness of the particles, or equivalently, the infinite speed of sound in the particle material. In the limit of infinite speed of sound all information on the exact collision mechanism, e.g., the sequence of individual pair collisions, is lost, since any of the collisions is of vanishing duration. Colliding particles of finite stiffness do not exhibit this feature, since here all processes take finite time.

According to these arguments the necessary information on the detailed collision mechanism is not accessible. The following set of assumptions turned out to yield realistic results, although they cannot be uniquely derived:
\begin{enumerate}
\item All individual pair collisions occur at once.
\item The transfer of momentum at contact points is finite. There is no momentum transfer which corresponds to attractive forces.
\item The relative velocity after a collision can never be smaller than $-\epsilon v$, where $\epsilon$ is the coefficient of restitution and $v$ is the impact velocity at that contact ($v<0$ !). 
\item If the velocity after the collision is strictly larger than $-\epsilon v$ there is no momentum transfer at this contact. 
\end{enumerate}
The first two assumptions have been introduced already. The remaining two assumptions deserve further discussion. Two-particle collisions can be described by means of the coefficient of restitution which relates the precollisional relative velocity $v$ and the final velocity $v^\prime$ after the collision: 
\begin{equation}
  v^\prime = -\epsilon v~.
\end{equation}
In the case of multi-particle collisions, however, two particles which initially rest relatively to each other may separate after a collision. Therefore, the final velocity may indeed be larger than the value $-\epsilon v$. Contrary it may no happen that two particle which collide with a finite impact velocity are at rest relative to each other afterwards. Therefore, $v^\prime \ge -\epsilon v$.

The fourth assumption simply means, that if two particles separate from each other with higher velocity than $-\epsilon v$ their aftercollisional velocity may not be increased further by an additional momentum transfer.

For convenience we define the excess velocity $\Delta v$
\begin{equation}
  \Delta v = v^\prime+\epsilon v\,,
\end{equation}
which is zero if $v^\prime=-\epsilon v$. The source of any velocity change is a momentum transfer between the colliding particles. We can relate the excess velocities at the contacts with the momentum transfers by means of the collisional geometry equation, which is derived in a similar way as the geometry equation of the force algorithm:
\begin{equation}
  \Delta v_\alpha = \left(1+\epsilon\right)v_\alpha + A_{\alpha\beta}^{\rm col} \Delta p_\beta\label{eq:geometrycollision}~.
\end{equation}
Note that the $v_\alpha$ are relative velocities of the contact points of two contacting particles, but not the velocities of the particles themselves. In mathematical terms the above discussed assumptions read
\begin{equation}
  \begin{split}
    \Delta v_\alpha&\ge0\\
    \Delta p_\alpha&\ge0\\
    \Delta v_\alpha\Delta p_\alpha &=0~.
  \end{split}
\end{equation}
The first condition prevents the final velocity from being smaller than $-\epsilon v$. The second condition excludes attractive interaction between particles. Finally, the third condition means that there may be a finite momentum transfer only if $\Delta v=0$. These conditions together with the geometry equation (\ref{eq:geometrycollision}) form a Linear Complementarity Problem which is already familiar from Sec. \ref{sec:Dantzig} and can be solved by Dantzig's Algorithm. 

\section{Resolution of static indeterminacy}\label{sec:indeterminacy}

If the number of contacts in the system is too large, the contact forces cannot be uniquely determined by the force algorithm. This occurs, if the number of free variables in the system, i.e. the number of contact forces, is larger than the number of mechanical degrees of freedom, $3N$ in 2d or $6N$ in 3d, with $N$ being the number of particles. However, the total forces and torques acting on the particles and, hence, their trajectories are unique. This drawback restricts the applicability of Rigid Body Dynamics for the simulation of railway ballast, since this system is of mainly static nature, the exact knowledge of the contact forces is crucial for understanding its behavior. 

\begin{figure}[htbp]
  \centerline{\psfig{figure=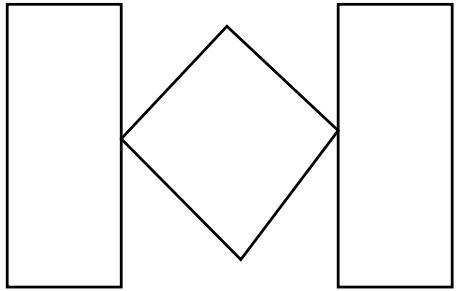,width=6cm}}
  \caption{The central particle is in contact with two other particles. The contact forces on both contacts are not independent of each other.}
  \label{fig:sillyexample}
\end{figure}

So far we have considered the contact forces as independent of each other. This assumption is the reason for the force indeterminacy. In realistic systems, however, the forces are not independent (Fig. \ref{fig:sillyexample}). If we apply an external force directed to the right on the central particle, we increase the contact force with the particle to its right while at the same time decreasing the contact force with the particle to its left. In this example both contact forces in reality depend on a single parameter, which is the applied external force. 

We can mimic this behavior by introducing small displacements of the particles. Each particle has a set of macroscopic coordinates $\vec{r}$ and $\vec{\phi}$ and a set of microscopic coordinates $\delta\vec{r}$ and $\delta\vec{\phi}$. In analogy to the vector $q$ of all (macroscopic) coordinates of all particles, we define the vector $\delta q$ of all microscopic (infinitesimal) coordinates. The contact network and the kinematic state of the particles is determined solely by the macroscopic coordinates, hence they can be considered as the actual coordinates of the particles. The microscopic coordinates lead to a deformation of contacting particles. By definition they are of infinitesimal size, which allows us to restrict ourselves to a linear approximation in $\delta q$ for the computation of the deformations. The vector $\xi$ of all deformations in the system
\begin{equation}
  \xi=\hat{D}\delta q
\end{equation}
is defined by the deformation matrix $\hat{D}$ and the microscopic coordinates. The dependence of $\hat{D}$ on the geometric properties of the systems is straightforward but lengthy, therefore we will not give explicit expressions here (for the full derivation see \cite{PoeschelSchwager:2002Algo}). We define a force law to relate the deformations at the contact points with the contact forces: 
\begin{equation}
  f=f(\xi)\,.
\label{eq:forcelaw}
\end{equation}
Now the contact forces are functions of the displacements $\delta q$. To calculate the forces we use the geometry equation (\ref{eq:geometrymodified}) together with the consistency conditions (\ref{eq:conditions5}). This set of equations is to be solved for the microscopic displacements $\delta q$. Hence, there are as many variables as degrees of freedom, i.e., the system is unique. 

To determine the microscopic coordinates we use an overdamped relaxation method. We start with a set of inconsistent coordinates $\delta q$. Inconsistent means that the consistency conditions for the resulting forces and accelerations are not fulfilled. Now we let the system relax. This way we find new microscopic coordinates such that $\xi^\prime=\xi-ha$, with $h$ being the step size and $a$ the vector of all contact accelerations. If the contact acceleration is negative the displacement will be larger, yielding a larger force to stop the approaching motion of the contacting particles. Since the adjustment step for the displacements is proportional to the acceleration the sequence of microscopic coordinates of the particles can be understood as overdamped motion. 

The adjusted microscopic coordinates are the solution of the linear system of equations 
\begin{equation}
  \xi-ha=\hat{D}\delta q^\prime
\end{equation}
or, equivalently
\begin{equation}
  \hat{D}\left(\delta q^\prime-\delta q\right)=-ha~.
\end{equation}
In cases where there are less degrees of freedom than contact accelerations the system of equations is overdetermined. There may be vectors $ha$ which are not representable by any vector $\delta q^\prime-\delta q$. In this case we have to project the vector $-ha$ into the image space of the operator $\hat{D}$ before solving the system of equations. 

We repeat the adjustment of microscopic coordinates until the consistency conditions are met. To further improve the speed of this method we can save the microscopic coordinates which yielded a consistent system in the previous time step. If the system did not change too much this set of microscopic coordinates is very close to the new solution and we need only few iteration steps to arrive at the new solution. 

This method combines the advantage of Rigid Body Dynamics, namely the ability to simulate very stiff particles, with the advantage of Molecular Dynamics, namely uniquely defined contact forces. An additional advantage of the method of small displacements is that we can now easily simulate certain degradation mechanisms. If we, for example, want to simulate the effects of abrasion of edges of the particles we can do this by gradually changing the force law (\ref{eq:forcelaw}), which describes changing edge properties of the particles. 

\section{Step size control}

The integration scheme used in our implementation is a Runge-Kutta method of fourth order. During one time step there are four force computations necessary. Since we use discrete time steps we are frequently faced with the problem that after a given time step some particles do deform each other. In this case, obviously, the chosen time step was too large. This problem could be solved by predicting the time of next contact (collision) from the present state of the particles. Since the particles are subject to forces which vary in time this prediction cannot be accurate, as we approach a collision we would have to update it repeatedly. Since this prediction method is quite complicated we chose a simpler method. We advance the system by a certain time step. When determining the contacts in the next time step (see Sec. \ref{sec:ablauf}) we check for deformations of the particles. If any deformations occur we restore the state of the system before this time step and take a time step of half its original value. We then repeat this process, i.e. advance the system by the new time step and check for deformations. If we again encounter deformations we again divide the time step by two and repeat the computation until a state without deformations is reached. The new state of the system is accepted and the other steps of the algorithm are performed. This procedure ensures that there will be no particle deformations in the system at the beginning of any accepted time step. 

Since this method can only decrease the time step, however, we need a procedure to increase the time step again\footnote{There is, of course, an upper limit to the time step which is dictated by the accuracy of the chosen integration scheme.} in order to avoid permanently slowing down the simulation. We can always increase the time step if there are no collisions in the systems for a certain time. Hence, if we successfully performed a number of time steps without encountering collisions we can increase the time step again by a factor of two. We cannot increase the time step after only one successful time step since this may lead to frequently alternating increase and decrease steps. A requirement of three successful time steps before increasing the time step has shown to yield good computational efficiency. 

With this choice of a step size control method we have completed the discussion of the simulation algorithm.

\section{Conclusions}

Although Molecular Dynamics methods have recently been applied successfully to the simulation of the dynamics of railway ballast \cite{Kruse:2002} we have doubts that this technique is suitable to the simulation of almost rigid, sharply edged particles such as railway ballast, in particular if the system behavior is governed mainly by static properties of the system. Moreover, it seems to be unsuitable for the simulation of long time effects such as densification and wear of ballast.

The Rigid Body Dynamics is a method which is intended to describe the motion of systems of very stiff particles. In a natural way it avoids the problems of Molecular Dynamics simulations which have been discussed in detail in Sec. \ref{sec:MDunsutable}. These problems are unavoidable within the concept of Molecular Dynamics. Therefore we believe that Rigid Body Dynamics is much better suited to the simulation of railway ballast than Molecular Dynamics. From the presentation of the algorithm we have seen that a time step in Rigid Body Dynamics is a lot more complicated than a time step in Molecular Dynamics. Hence the according implementation requires by far more computing time than an implementation of a Molecular Dynamics code. Fortunately this disadvantage is offset by the fact that in Rigid Body Dynamics we can chose a larger time step than in Molecular Dynamics. Whereas in Molecular Dynamics the time step is determined by the critical deformation (the length on which the force changes by one unit) divided by the characteristic velocity in the system, which usually yields a very small time step (e.g. $\sim10^{-7}$ sec), the time step in Rigid Body Dynamics is determined by the characteristic time in which the forces in the system change significantly. The latter quantity is usually much larger ($\sim10^{-3}$ sec). Thus, the computing time requirements for both methods are comparable when simulating mainly static problems as it is the case for railway ballast. 

An additional advantage of Rigid Body Dynamics is that fracture of particle can be introduced in a very natural way. When dealing with deformable particles in MD simulations the fracture of particles can lead to numerical difficulties. After the fracture of a particle, i.e., when a large particle is replaced by two or more fragments, there will be a finite time in which the system has to readjust and to reach a new stable state. There are situations when the new stable state deviates significantly from the realistic (experimental) situation. This problem is not encountered in Rigid Body Dynamics. After a fragmentation (which is algorithmically done by constructing two or more new polygons from the previous particle simply by introducing another edge into the old particle and splitting along this line) the system remains always stable. 

It is our firm believe that Rigid Body Dynamics provides a very promising alternative to Molecular Dynamics for the simulation of railway ballast.

\end{document}